\input phyzzx\def\e{\adveq\eqno{\rm (\chapterlabel.\the\equanumber)}}
\def\mysec#1{\equanumber=0\chapter{#1}}
\def\adveq{\global\advance\equanumber by 1}
\def\myeq{{\rm \chapterlabel.\the\equanumber}}
\def\rarrow{\rightarrow}

\def\twoline#1#2{\displaylines{\qquad#1\hfill(\adveq\myeq)\cr\hfill#2
\qquad\cr}}

\def\semidirect{\mathrel{\raise0.04cm\hbox{${\scriptscriptstyle |\!}$
\hskip-0.175cm}\times}}


\def\ref#1{$^{[#1]}$}

\def\r#1{$[\rm#1]$}

\def\threeline#1#2#3{\displaylines{\qquad#1\hfill\cr\hfill#2\hfill\llap{(\adveq\myeq)}\cr
\hfill#3\qquad\cr}}

\def\e{\adveq\eqno{\rm (\chapterlabel.\the\equanumber)}}
\def\mysec#1{\equanumber=0\chapter{#1}}
\def\adveq{\global\advance\equanumber by 1}
\def\myeq{{\rm \chapterlabel.\the\equanumber}}
\def\rarrow{\rightarrow}

\def\twoline#1#2{\displaylines{\qquad#1\hfill(\adveq\myeq)\cr\hfill#2
\qquad\cr}}

\def\semidirect{\mathrel{\raise0.04cm\hbox{${\scriptscriptstyle |\!}$
\hskip-0.175cm}\times}}


\def\ref#1{$^{[#1]}$}

\def\r#1{$[\rm#1]$}

\def\threeline#1#2#3{\displaylines{\qquad#1\hfill\cr\hfill#2\hfill\llap{(\adveq\myeq)}\cr
\hfill#3\qquad\cr}}

\def\half{{1\over2}}

\overfullrule=0pt
\date{May, 2020}
\date{May, 2020}
\titlepage
\title{$B_k$ Spin Vertex Models and Quantum Algebras}
\author{Doron Gepner}
\vskip20pt
\line{\it\hfill  Department of Particle Physics and Astrophysics, Weizmann Institute,\hfill}
 \line{\it\hfill Rehovot 76100,  Israel\hfill} 
 
 \abstract
 We construct new solvable vertex models based on the spin representation of the Lie algebra $B_k$.
 We use these models to study the algebraic structure underlying such vertex theories. We show
 that all the $B_k$ spin vertex models obey a version of the BMW algebra along with extra relations that are
 called $n$--CB (conformal braiding) algebras. These algebras were discussed before for
 various IRF (interaction round the face) models. Here we establish that the same algebras
 hold for vertex models.
 
 \endpage

\mysec{Introduction.}
Solvable lattice models in two dimensions are an excellent playing ground to study 
phase transitions, integrable models and knot theory. For reviews see
\REF\Baxter{R.J. Baxter, ``Exactly solved models in statistical mechanics'', Academic Press, London,
England, 1982.}
\REF\Wadati{M. Wadati, T. Deguchi and Y. Akutsu, Physics Reports 180 (4) (1989) 247.}
\r{\Baxter,\Wadati}.

Of particular significance to us is the algebraic structure underlying solvable lattice models.
Examples of  such algebras are the Temperley--Lieb algebra
\REF\TL{N. Temperley and E. Lieb, Proc. R. Soc. A 322 (1971) 251.}
\r\TL\
and the Birman--Murakami--Wenzl algebra (BMW)
\REF\BW{J.S. Birman and H. Wenzl, Trans. Am. Math. Soc. 313 (1) (1989) 313.}
\REF\Mur{J. Murakami, Osaka J. Math. 24 (4) (1987) 745.}
\r{\BW,\Mur}. These algebras had a major role in the solution of the models as well
as applications such as knot theory. In particular, in knot theory we mention the celebrated work of Jones
\REF\Jones{V.F.R. Jones, Int. J. Mod. Phys. A5 (1990) 441.}
\r\Jones\
and the works of Wadati et al. reviewed in ref. \r\Wadati. See also the book
\REF\Kauf{L.H. Kauffman, ``Knots and physics", World Scientific, Singapore (1991).}
\r\Kauf.

In recent works
\REF\CBtwo{V. Belavin and D. Gepner, Nucl. Phys. B 938 (2019) 223.}
\REF\CBthree{V. Belavin and D. Gepner, JHEP 02 (2019) 033.}
\REF\CBfour{V. Belavin, D. Gepner, J.R. Li and R. Tessler, JHEP 11 (2019) 155.}
\REF\CBfive{V. Belavin and D. Gepner, arXiv: 2001.09280 (2020).}
\r{\CBtwo,\CBthree,\CBfour,\CBfive},
the algebraic structure of IRF solvable lattice models was established.
These works
were based on the Yang--Baxter equation and the ansatz for Baxterization put forward
in ref.
\REF\Found{D. Gepner, arXiv: hep-th/9211100v2 (1992).}
\r\Found, generalizing the two blocks Baxterization of Jones \r\Jones, to more than two blocks.
An algebra was described for any number of blocks and called the $n$--CB algebra (conformal braiding),
where $n$ is the number of blocks (defined as the order of the polynomial equation satisfied by the Boltzmann weights). The $n$--CB algebra includes the Temperley--Lieb algebra
and a version of BMW algebra along with additional relations.

\REF\Arb{J.H. Arbeitman, S. Mantilla and I. Sodeman, Phys. Rev. B 99 (2019) 245108.}
\REF\Bossard{A. Bossard and W. Galleas, J. of Math. Phys. 60 (2019) 103509.}
\REF\Brubker{B. Brubaker, V. Buciumas, D. Bump and H. Gustafsson, arXiv: 1902.01795v3 (2019).}
\REF\Nirov{K.S. Nirov and A.V. Razumov, SIGMA 15 (2019) 068.}

Our aim here is to study the $n$--CB algebra for vertex models. 
For recent works on vertex models, see \r{\Arb,\Bossard,\Brubker,\Nirov}.
We establish that the same algebra
is obeyed by vertex models. For this study, we construct new vertex models based on the spin 
representation of the simple Lie algebra $SO(2k+1)$ which is denoted as $B_k$, for arbitrary 
positive integer $k$.
These models are described for any number of blocks which is $n=k+1$.
Previously, a Yang--Baxter solution was described for the vector representation of $B_k$  (for a review see \r\Wadati\ and refs. therein.)

\mysec{Vertex models and their Baxterization.}
\par

Vertex lattice models are described by an element of End$(V\otimes V)$ where $V$ is some vector
space. It will be convenient to describe these elements using a matrix notation. Namely, if 
$R\in {\rm End}(V\otimes V)$ then
we may write,
$$R(v_\mu\otimes v_{\nu})=R_{\mu,\nu}^{\bar \mu,\bar\nu} (v_{\bar \mu}\otimes v_{\bar \nu}),\e$$
where $\nu$ and $\mu$ are basis vectors of the vector space $V$. Here,
we include the indices of $R$.

The matrix $R$ depends on the spectral parameter $R(u)$ where $u$ is some complex number.
The solvability of the model is encapsulated in the Yang--Baxter equation (YBE) which can be
written as an equation in End$(V\otimes V\otimes V)$,
$$(R(u)\otimes 1) (1\otimes R(u+v)) (R(v)\otimes 1)=(1\otimes R(v)) (R(u+v)\otimes 1) (1\otimes R(u)).\e$$
This equation can be expanded in terms of matrix elements, eq. (2.1), to give,
$$\sum_{\alpha,\beta,\gamma} R_{j,k}^{\beta,\alpha}(u) R_{i,\beta}^{l,\gamma} (u+v) R_{\gamma,\alpha}^{m,n} (v)=
\sum_{\alpha,\beta,\gamma} R_{i,j}^{\alpha,\beta} (v) R_{\beta,k}^{\gamma,n} (u+v) R_{\alpha,\gamma}^{l,m}(u).\e$$
We assume that $R(u)$ is a trigonometric solution of the Yang--Baxter equation.

The vertex models may obey a number of properties in addition to the YBE.
The initial condition,
$$R_{i,j}^{k,l}(0)=\delta_{ik} \delta_{jl}.\e$$
The inversion relation,
$$\sum_{m,n} R_{i,j}^{m,n}(u) R_{m,n}^{l,k}(-u)=\rho(u) \rho(-u) \delta_{il} \delta_{jk},\e$$ 
where $\rho(u)$ is a function, to be specified later.
Also, crossing symmetry,
$$R_{ji}^{k,l}(u)=R^{l,\bar i}_{\bar k, j}(\lambda-u) \left[ {r(i) r(l)\over r(j) r(k)}\right]^{1/2},\e$$
where $\lambda$ is the crossing parameter. Here $\bar i$ is 
the charge conjugation of $i$ and the crossing multiplier is $r(i)$, where $r(\bar i)=1/r(i)$. Usually, in a vector model, we will 
have that $\bar v=-v$.
We have the reflection symmetry,
$$R_{i,j}^{m,n} (u)=R_{m,n}^{i,j} (u).\e$$
Finally, we have the charge conservation,
$$R_{m,n}^{i,j}=0,\quad{\rm unless\ } m+n=i+j.\e$$

We find it convenient to define an operator form for the $R$ matrix. We define the matrix,
$$X_i(u)=\sum_{m,n,a,b} R^{a,b}_{m,n}(u) I^{(1)} \otimes \ldots \otimes I^{(i-1)}\otimes 
e^{(i)}_{a,m} \otimes e_{b,n}^{(i+1)} \otimes I^{(i+2)} \otimes \ldots \otimes I^{(n)},\e$$
where $\otimes$ means tensor product, $I^{(i)}$ is the identity matrix at position $i$,
and $e_{ab}$ is a matrix whose elements are given by $(e_{rs})_{lm}=\delta_{rl} \delta_{sm}$.
We define in a similar fashion other operators.
It is then clear that the YBE, eq. (2.2), can be written as
$$X_i(u) X_j(v)=X_j(v) X_i(u),\quad {\rm if\ }|i-j|\geq 2,$$
$$X_i(u) X_{i+1}(u+v) X_i(v)=X_{i+1}(v) X_i(u+v) X_{i+1} (u).\e$$

We will build the vertex model from the data of a fixed conformal field theory. Given the conformal
filed theory $\cal O$, let $V$ be the representation of some primary field in $\cal O$. The vertex model is given in terms of the representations that appear in the tensor product of $V$.
We assume that the theory is real and that $[V]$ is a real representation.
Thus, we have the fusion product,
$$[V]\times [V]=\sum_{i=0}^{n-1} [\psi_i],\e$$
where $[x]$ denotes the primary field $x$. Here $n$ is an integer which is called the number of blocks
and $[\psi_0]=1$, is the unit representation. The order of the fields in eq. (2.11) is important as will be discussed later. For real models a rule of thumb appears for this order. Namely,  the field
$\psi_{i+1}$ appears in the tensor product of $\psi_i$ and the adjoint representation
(for quantum groups). Thus, in particular, $\psi_0=1$ and $\psi_1$ is the adjoint representation.
The complete implementation of this rule is presently not clear.
For each representation that appears in this fusion product we define a projection operator $P_i$
onto this representation. For this purpose, we define the limit of the trigonometric solution of 
the Yang--Baxter equation, $X_i(u)$,  as,
$$X_i=\lim_{u\rarrow i\infty} e^{i(n-1) u} X_i(u),\qquad X_i^t=\lim_{u\rarrow -i\infty} e^{-i(n-1) u} X_i(u).\e
$$  

The eigenvalues of $X_i$ can be seen from conformal field theory to be,
$$\lambda_i=\epsilon_i e^{i\pi(2\Delta_v-\Delta_i)},\e$$
where $\Delta_v$ is the conformal dimension of the primary field $[V]$, $\Delta_i$ is the conformal dimension of the representation $[\psi_i]$ and $\epsilon_i=\pm1$ indicating whether the product is
symmetric or antisymmetric.

From $X_i$ we can define the $a$th projection operator as,
$$P_i^a=\prod_{p\neq a} \left[ {X_i-\lambda_p I \over \lambda_a-\lambda_p  }\right],\e$$
where $a=0,1,\ldots,n-1$ and  $I$ is the unit operator. We have the following relations for the projection operators,
$$X_i=\sum_{a=0}^{n-1} \lambda_a P_i^a,\e$$
$$\sum_{a=0}^{n-1} P_i^a=I,\qquad P_i^a P_i^b=\delta_{ab} P_i^a.\e$$

From the projection operator one may build the solution to the YBE, $X_i(u)$.
It is basically the same conjecture as for the IRF models described in ref. \r\Found. We define the parameters by,
$$\zeta_i=\pi (\Delta_{i+1}-\Delta_i)/2,\e$$
and $\lambda=\zeta_0$ is the crossing parameter and $i=0,1,\ldots,n-2$.
The trigonometric solution to the Yang--Baxter equation ansatz is then,
$$X_i(u)=\sum_{a=0}^{n-1} f_a(u) P_i^a,\e$$
where the functions $f_a(u)$ are given by,
$$f_a(u)=\left[ \prod_{r=1}^a \sin(\zeta_{r-1}-u) \right ] \left[ \prod_{r=a+1}^{n-1} \sin(\zeta_{r-1}+u)\right]\bigg/
\left[ \prod_{r=1}^{n-1} \sin(\zeta_{r-1})\right].\e$$

From the ansatz it is easy to see that the inversion relation, eq. (2.5),  holds with
$$\rho(u)=\prod_{r=1}^{n-1} {\sin(\zeta_{r-1}-u)\over \sin(\zeta_{r-1})}.\e$$
The crossing equation, eq. (2.6), holds with the crossing parameter $\lambda=\zeta_0$.

The order of the fields $[\psi_i]$ is important and the YBE holds only for one particular order.
We will specify below the order which is suitable for specific models.

An interesting question is the relation between the CFT used to define the model and the conformal
field theories which arise at the criticality of the models. A partial answer, based on 
(D. Gepner, unpublished work), is that the critical field theories are cosets of the original theory,
where in one limit, the original theory is in the denominator, whereas in the other it is in the numerator
of the coset theory. The full coset theory is presently unknown, except in some examples.
For a review see ref. \r\Wadati.

We are interested in the algebra obeyed by these models. For this purpose, we define
the operators,
$$G_i=2^{n-1} e^{-i(n-1)\zeta_0/2} \left [\prod_{r=1}^{n-1} \sin(\zeta_{r-1})\right] X_i,\e$$
and
$$G_i^{-1}=2^{n-1} e^{i(n-1)\zeta_0/2} \left[\prod_{r=1}^{n-1}\sin(\zeta_{r-1})\right]  X_i^t,\e$$
where $X_i$ and $X_i^t$ are given by eq. (2.12).
We also define the operator,
$$E_i=X_i(\lambda).\e$$
The inversion relation eqs. (2.5, 2.20) implies that as defined $G_i G_i^{-1}=I$.

From the crossing relation, eq. (2.6), it follows that $E_i$ can be expressed as follows,
$$E_{a,b}^{m,n}=\delta_{\bar a, b} \delta_{\bar m, n} r(n) r(b),\e$$
where $r(a)$ is the crossing multiplier. Here we reverted back to the explicit notation for $E_i$.
From the above equation, it follows that $E_i$ obeys the relation,
$$E_i E_{i\pm1} E_i=E_i,\e$$
where we used the equation $r(\bar m)=1/r(m)$.
From the ansatz eqs. (2.18,  2.19) it follows that 
$$E_i^2=b E_i,\e$$
where 
$$b=\prod_{r=1}^{n-1} {\sin (\zeta_0+\zeta_{r-1})\over \sin(\zeta_{r-1})}.\e$$
These two equations together are the celebrated Temperley--Lieb algebra \r\TL. Thus, we proved that 
any real vertex model, with any number of blocks, obeys the Temperley--Lieb algebra,
assuming that the ansatz eq. (2.18-2.19) holds.

Since $E_i$ is proportional to $P_i^0$ we find the following relations,
$$G_i E_i=E_i G_i=l^{-1} E_i,\e$$
where $l$ is given by,
$$l=i^{n-1} \exp\left[i(n-1)\zeta_0/2+i\sum_{r=0}^{n-2} \zeta_r\right].\e$$

From the YBE, eq. (2.2), we find that $G_i$ obeys the braid group relation,
$$G_i G_j=G_j G_i \quad {\rm if\  } |i-j|\geq2,\qquad  G_i G_{i+1} G_i=G_{i+1} G_i G_{i+1}.\e$$

From the ansatz, eqs. (2.18,2.19), and from the equation $\sum_a P_i^a=I$ we find the skein relation,
$$G_i^{n-2}=a E_i +\sum_{r=-1}^{n-3} b_r G_i^{r},\e$$
where the coefficients $a$ and  $b_r$ are expressed as functions of the parameters $\zeta_i$,
which can be calculated from the ansatz, eqs. (2.18, 2.19).

\mysec{Vertex models and quantum groups.}

We utilize now the powerful method for constructing solutions to the YBE vertex models, eq. (2.2),
which is quantum groups
\REF\Jimbo{M. Jimbo, Lett. in Math. Phys. 10 (1985) 63.}
\REF\Drinfeld{V.G. Drinfeld, Doklady Akad. Nauk. SSSR 283 (5)  (1985) 1060.}
\REF\Pasquier{V. Pasquier, Comm. Math. Phys. 118 (1988) 355.}
\r{\Jimbo,\Drinfeld,\Pasquier}.

The definition of the quantum group is as follows \r{\Jimbo,\Drinfeld}. Let $A=(a_{ij})$ be a
Cartan matrix of a simple Lie algebra $G$. Let $\{\alpha_j\}$ and $\{h_j\}$ be the simple roots and coroots,
for $1\leq i\leq N$, 
such that $<h_i | \alpha_j>=a_{ij}$. For a parameter $q$ which is nonzero we define $q_i=q^{(\alpha_i,\alpha_i)}$, where $(|)$ is the invariant inner product in $h^*$.

The generators of the quantum group are $\{ k_i^{\pm1}, e_i,f_i\}_{1\leq i\leq N}$.  They obey the relations,
$$k_i k_i^{-1}=k_i^{-1} k_i=1,\qquad [k_i,k_j]=0,\e$$
$$k_i e_j k_i^{-1}=q_i^{a_{ij}}e_j,\qquad k_i f_j k_i^{-1}=q_i^{-a_{ij}} f_j,\e$$
$$[e_i,f_j]=\delta_{ij} (k_i^2-k_i^{-2})/(q_i^2-q_i^{-2}).\e$$
There are additional relations, (3D) and (3E) of ref. \r\Jimbo, but we will not require these.

For $q\rarrow 1$ the quantum algebra, denoted by $U_q(G)$ reduces to the simple Lie algebra $G$.
(Actually, the quantum group can be defined for any Kac--Moody algebra \r{\Jimbo,\Drinfeld}.)

We shall need the co--product of the quantum group $U_q(G)$. This is the homomorphism
$\Delta^{(m)} U\rarrow \otimes^{m} U$  ($m$ fold tensor product), defined by
$$\Delta^{(m)}(k_i)=k_i\otimes k_i\otimes \ldots \otimes k_i,\e$$
$$\Delta^{(m)}(X_i)=\sum_{\nu=1}^ m k_i\otimes\ldots \otimes k_i \otimes^\nu  X_i\otimes k_i^{-1}\otimes\ldots\otimes 
k_i^{-1},\e$$
for $X_i=e_i$ or $X_i=f_i$.  The co--product obey the same quantum group $U_q(G)$.

In the following we will assume that $q$ is not a root of unity, unless otherwise specified. 
In this case, the irreducible representations
of $U_q(G)$ are labeled by the irreducible representations of $G$ and have the same dimensions. 

The solution to the vertex YBE, eq. (2.2), commutes with the co--product,
$$[R,\Delta^{(2)}(X_i)]=0,\e$$
for any $X_i=e_i$ or $f_i$ or $k_i$. This equation is not enough to determine the $R$ matrix.
However, given a solution to this equation, it is guaranteed to have the same eigenvectors but 
not the same eigenvalues as the $R$ matrix. Thus, they share the same projection operators,
eq. (2.14). Assuming that the number of distinct eigenvalues of $R$ is $n$, where $n$ is the number of  
blocks, than the projection operators are given as in eq. (2.14),
$$P^a=\prod_{p\neq a} \left[ {R-\lambda_p I \over \lambda_a-\lambda_p}\right],\e$$
where $\lambda_p$ are the eigenvalues of $R$.
We can than use our ansatz eqs. (2.18, 2.19) to get the full trigonometric solution of the YBE.

Each projection operator $P^a$ is associated to some representation in the tensor product $g\in V\times V$, where
$V$ is the representation used to define the vertex model. The projection operator can be written
as,
$$(P^g)_{a,b}^{c,d}=\sum_\lambda <g \,\lambda | V\, a\, V\, b><g\,\lambda | V\,c\,V\,d>,\e$$
where $\lambda$ runs over the weights of the representation $g$ and $<g\,\lambda | V\,a\, V\,b>$
is the quantum group Wigner coefficient of this tensor product. $P^g$ is the vertex projection operator with 
the weights $a,b,c,d$ which are weights of  the representation $V$. From this equation, it is clear that the projection operator vanishes unless
$a+b=c+d$, eq. (2.8). For $SU(2)$ this was described in ref. \r\Pasquier.
In this reference, it was shown that for $SU(2)$ the vertex and the IRF models have the same
Baxterization.

\mysec{$B_k$ spin vertex models.}

Our purpose is to describe solvable vertex models based on the algebra $B_k$, or $SO(2 k+1)$,
where the representation $V$ is the spinor representation. We use the basis for $B_k$ where 
the simple roots are $\alpha_n=\epsilon_{n}-\epsilon_{n+1}$, for $n=1,2,\ldots,k-1$ and
$\alpha_k=\epsilon_k$. Here $\epsilon_i$ are orthogonal unit vectors. The spinor representation
has the highest weight $(\epsilon_1+\epsilon_2+\ldots \epsilon_k)/2$ and the weights of
this representation are $(\pm\epsilon_1\pm\epsilon_2\pm\ldots\pm \epsilon_k)/2$. 
We find it 
convenient to add $1/2$ to these weights and to represent the weights of the spinor representation
by $m$ where $m_i=0 \ {\rm or\ } 1$.

We look for a solution $C$ for the spinor representation of the algebra $B_k$, 
which commutes with the co--product, eq. (3.6). 
Such a solution was described recently in a paper by Wenzl
\REF\Wenzl{H. Wenzl, ``Dualities for spin representation", arXiv: 2005.11299, (2020).} \r\Wenzl.
The solution $C$ is an element of End$(V\otimes V)$ where $V$ denotes the spinor representation.
It is given by \r\Wenzl,
$$\twoline{C_{m,n}^{b,c}=\sum_{j=1}^k \delta_{m_j,1-n_j} (-q^2)^{\{m-n\}_j} \delta_{b,\bar m^j} \delta_{c,\bar n^j}
+}{
(-1)^k \delta_{m,b} \delta_{n,c}(-q^2)^{\{m-n\}_k}/[2],}$$
where 
$$\{ m\}_j=\sum_{r=1}^j m_r,\e$$
and
$\bar n_j$ is equal to $n$ except at the $j$th coordinate where it is  $1-n_j$. Here [2]=$q+q^{-1}$.
Here $m,n,b,c=0 {\rm \  or\ 1}$ are weights of the spinor representation shifted by $1/2$.
The matrix $C$, so constructed, commutes with the co--product eq. (3.6). 

The eigenvalues of the matrix $C$ were computed by Wenzl \r\Wenzl. They are given by
$$\lambda_j=(-1)^j s(k+\half-j),\quad {\rm for\ } j=0,1,\ldots ,k,\e$$
where 
$$s(x)={q^{2 x}-q^{-2 x} \over q^2-q^{-2}}.\e$$

There are $k+1$ distinct eigenvalues of $C$. 
Thus, this is a $k+1$ blocks theory.
Each eigenvalue corresponds to a representation
in the tensor product $V\times V$, where $V$ is the spinor representation. The $j$th eigenvalue
$\lambda_j$ corresponds to the representation $V_j=\wedge^j v$ where $v$ is the vector representation.
The highest weight of the representation $V_j$ is $\epsilon_1+\epsilon_2+\ldots+\epsilon_j$. It is the fully anti--symmetric
representation in the tensor of $j$ vector representations.

We assume that $q$ is not a root of unity and is nonzero. To connect with section (3),
we identify
$$q^2=\exp[\pi i/(r+g)],\e$$
where $r$ is the level of the WZW model based on $B_k$, at level $r$ and $g$ is the
dual Coxeter number,
$$g=2 k-1.\e$$
We assume that the level $r$ is not a real rational number, so that $q$ is not a root of unity.
The dimension of the representation with highest weight $\Lambda$ in a WZW theory is given by
$$\Delta_\Lambda={\Lambda(\Lambda+2 \rho) \over 2(r+g)}.\e$$
Here $\rho$
is half the sum of positive roots and $C_\Lambda=\Lambda(\Lambda+2 \rho)$ is the Casimir
of the representation $\Lambda$. See, e.g. 
\REF\Francesco{P. Francesco, P. Mathieu and D. Senechal, ``Conformal field theory", Springer.}
\r\Francesco. 

As explained in section (3), the eigenvectors of $C$ are the projections of the solution of the YBE
to the different representations. We thus define,
$$(P^a)_{m,n}^{b,c}=\prod_{p\neq a} \left[ {C-\lambda_p I\over \lambda_a-\lambda_p}\right],\e$$
where the product is in End$(V\otimes V)$ and $I$ is the identity map.

We know from equation (2.13)  that the eigenvalues of the $R$ matrix are given by $\epsilon_j\exp[-i\pi\Delta_j]$
up to an irrelevant constant. Thus, we need to compute the second Casimir of the representations $V_j$, since the dimensions of the representations are computed in terms of the Casimir, eq. (4.7).
The Casimir is given by
$$ C(V_j)= C_j=j(2k+1-j).\e$$
Thus the eigenvalues of $R$ are 
$$\beta_j=\epsilon_j q^{-C(V_j)},\e$$
where $\epsilon_j$ is a sign which is harder to compute.
To give this sign we define,
$$(h_0,h_1,\ldots ,h_k)=(0,2,4,\ldots ,k, k-1,k-3,\ldots,1),\e$$
for even $k$. For odd $k$,
$$(h_0,h_1,\ldots ,h_k)=(0,2,\ldots,k-1,k,k-2,k-4,\ldots,1).\e$$
Then the sign $\epsilon_j$ is given by,
$$\epsilon_{h_s}=(-1)^s.\e$$

We are now in position to construct the $R$ matrix as
$$R_{m,n}^{a,b} =\sum_{j=0}^k \beta_j (P^j)_{m,n}^{a,b},\e$$
This is since we know the projection operators from eq. (4.8) and the eigenvalues of $R$ from eq. (4.10).

We can now check that the $R$ matrix, so constructed, obeys the braiding relation,
 $$\sum_{\alpha,\beta,\gamma} R_{j,k}^{\beta,\alpha} R_{i,\beta}^{l,\gamma}  R_{\gamma,\alpha}^{m,n} =
\sum_{\alpha,\beta,\gamma} R_{i,j}^{\alpha,\beta}  R_{\beta,k}^{\gamma,n}  R_{\alpha,\gamma}^{l,m}.\e$$
We checked this $R$ matrix numerically for $k=2,3,4,5,6$ and it is, indeed, obeyed for various
weights and for general $q$.

We can now build the full solution to the YBE, eq. (2.3). We need to compute the parameters 
$\zeta_i$. To do this, we need to know the order of the operators $\psi_i$ in eq. (2.11). In fact, the order 
of the representations is given by $h_r$. Thus, we have
$$\zeta_j=(C_{h_{j+1}}-C_{h_j})/2,\e$$
for $j=0,1,\ldots,k-1$.
We also replace the $\sin(x)$ in eq. (2.19) by
$$\sin(x)\rarrow p(x)=q^x-q^{-x}.\e$$
Then the solution to the YBE, eq. (2.3), assumes the form,
$$X_{m,n}^{a,b} (u)=\sum_{j=0}^{k} f_j(u) (P^{h_j})_{m,n}^{a,b} ,\e$$
where
$$f_a(u)=\left[ \prod_{j=1}^a p(\zeta_{j-1}-u) \right ] \left[ \prod_{j=a+1}^{k} p(\zeta_{j-1}+u)\right]\bigg/
\left[ \prod_{j=1}^{k} p(\zeta_{j-1})\right],\e$$
where $a=0,1,\ldots,k$.

For example for $k=6$ we have, $(\zeta_0,\zeta_1,\ldots,\zeta_5)=(11,7,3,-1,-5,-9)$. The crossing
parameter is always $\lambda=\zeta_0$.

We can now check numerically that the matrix $X_i(u)$ so defined obeys the Yang--Baxter equation,
eq. (2.3). We checked this numerically for $k=2,3,4,5,6$ for various values of the weights and the spectral parameters
and indeed the YBE is obeyed for general $q$.

Actually, our results holds also for $q$ which is a root of unity. We take $q^2=\exp[i\pi s /(r+g)]$, as
in eq. (4.5), where $r$ and $s$ are now integers such that, $\gcd(s,r+g)=1$. Then, if the level $r$
is greater or equal two, then the fusion rule in eq. (2.11) is the same as the tensor product,
since the representations $\psi_i$ appear at level two.
Namely, all the representations appear if the level is greater or equal two. Thus, the ansatz eqs. (4.18, 4.19),
holds as it is. We  checked this for various algebras of the type $B_k$ and various integer levels, $r$,
and indeed the YBE is obeyed for $q$ which is  a root of unity, as well.
Thus, for levels greater than one, exactly the same solution holds. We call these models
for rational level,
the restricted models.

\mysec{$n$--CB algebra and $B_k$ vertex models.}

The $B_k$ vertex models are $k+1$ blocks models. For $k=2$ this is a three blocks model.
Thus, it is natural that the model would obey the BMW algebra \r{\BW,\Mur}, as we will show.
We use the operator notation eq. (2.9) and define the operators $G_i$ and $E_i$ as before,
eqs. (2.21-2.23).
The relations of the BMW algebra are,
$$G_i-G_i^{-1}=m(1-E_i),\e$$
$$G_i G_j=G_j G_i\ {\ \rm if\ } |i-j|\geq 2,\qquad  G_iG_{i+1} G_i=G_{i+1} G_i G_{i+1},\e$$
$$E_i E_{i\pm1} E_i=E_i,\qquad E_i^2=b E_i,\e$$
$$G_{i\pm1} G_i E_{i\pm1}=E_i G_{i\pm1} G_i=E_i E_{i\pm1},\qquad G_{i\pm1} E_i G_{i\pm1}=
G_i^{-1} E_{i\pm1} G_i^{-1},\e$$
$$G_{i\pm1} E_i E_{i\pm1}=G_i^{-1} E_{i\pm1},\qquad E_{i\pm1} E_i G_{i\pm1}=E_{i\pm1} G_i^{-1},\e$$
$$G_i E_i=E_i G_i=l^{-1} E_i,\e$$
$$E_i G_{i\pm1} E_i=l E_i,\qquad E_i G_{i\pm1}^{-1} E_i=l^{-1} E_i,\e$$
where 
$$b=m^{-1}(l-l^{-1})+1,\e$$
and $l$ and $b$ are given by eqs. (2.27, 2.29) for three blocks, $n=3$.

We checked the BMW relations eqs. (5.1-5.8) for the $B_2$ vertex model and indeed they are all obeyed
for various weights and general $q$. We find,
$$b=-(q^4+q^2+q^{-2}+q^{-4}),\e$$
and
$$l=-q^5,\qquad m=q+q^{-1} \e$$

In fact, as we show,  the BMW algebra is also obeyed for $k>2$, except for the skein relation, eq. (2.31).
The relations eqs. (5.2, 5.3, 5.6) were already proved in section (2) for all the vertex models, eqs. 
(2.25, 2.26, 2.28, 2.30),
along with the new skein relation eq. (2.31).
It remains to check the other relations. We checked them for $k=3,4,5,6$ with various weights,
and general $q$, and indeed they are all obeyed. The parameters $l$ and $b$ are given by eqs. (2.27, 2.29).
We call this algebra BMW$^\prime$.

We checked the BMW$^\prime$ algebra also for the restricted models and it also holds. Our discussion
below applies equally well to the restricted models as they also obey the same ansatz and the
same YBE.

In ref. \r{\CBtwo,\CBthree,\CBfour,\CBfive}, we analyzed the Yang--Baxter equation assuming only
BMW$^\prime$ algebra and the ansatz eqs. (2.18, 2.19). We established this only for IRF models and not
for vertex models. However, all the assumptions are exactly the same, even though the definition of
the operators is different, eq. (2.9). Thus, the same conclusions we found by expanding the YBE still
hold.
We found that for three blocks, $k=2$, we get a week version of the BMW algebra \r\CBfour.  For four blocks, $n=4$, we get an algebra which we called $4$--CB, which is BMW$^\prime$, along
with one additional relation. The additional relations are enlisted in the appendix.
For five blocks $n=5$ we get additional $19$ relations which are quite 
bulky. This method can be used to compute the algebra for any number of blocks, $n$, which we call
$n$-CB algebra.

Since all of the assumptions are the same for IRF models as for the vertex models, we conclude that
the $n$-CB algebra holds for the $B_k$ vertex models, with $n=k+1$. We conjecture that the
$n$--CB algebra is obeyed for all the solvable vertex models with $n$ blocks, for which the ansatz
eqs. (2.18, 2.19) holds.

\appendix

\line{\bf Four CB relations.\hfill}

For completeness, we summarize here the four blocks relations  \r\CBfour. The skein relation is given by
\def\frac#1#2{{#1\over #2}}
$$\threeline{
G_i^2= i e^{-\frac{1}{2} i \zeta_0-i \zeta_1-i \zeta_2} \left(1-e^{2 i \zeta_1}+e^{2 i \zeta_1+2 i \zeta_2}\right) \
G_i+i e^{-\frac{3}{2} i \zeta_0+i \zeta_1-i \zeta_2} \
G_i^{-1}}{+\frac{e^{-2 i \zeta_0-2 i \zeta_1-2 i \zeta_2} \left(e^{2 i 
\zeta_1}-1\right) \left(1+e^{2 i \zeta_0+2 
i \zeta_1+2 i \zeta_2}\right) \left(e^{2 i \zeta_2}-1\right) 
 }{\left(e^{2 i \zeta_0+ 2 i \zeta_2}-1\right) }E_i}{-e^{-i \zeta_0-2 i 
\zeta_2} \left(1-e^{2 i \zeta_2}+e^{2 i \zeta_1+2 i \zeta_2}\right).}$$

The last relation follows from the Yang Baxter equation and the ansatz eq. (2.18-2.19). It is 
$$g(i,i+1,i)=g(i+1,i,i+1),\e$$
where 
$$\threeline{
g=a_{1,2,4}+a_{1,3,1}+a_{4,2,1}+i q^{-\zeta_0/2+\zeta_1-\zeta_2} (a_{1,3,4}+a_{4,2,4}+a_{4,3,1})+}{
i q^{\zeta_0/2-\zeta_1+\zeta_2}  (a_{2,3,4}+a_{4,1,4}+a_{4,3,2})+}{
i {q^{\zeta_1+\zeta_2}\over (q^{2\zeta_1}-1)(q^{2\zeta_2}-1)}\left(q^{\zeta_0/2} a_{1,2,1}+q^{-\zeta_0/2} a_{2,1,2} \right)+z a_{4,3,4},}$$
where 
$$\twoline{z={q^{-\zeta_0-2\zeta_1-2\zeta_2} (q^{2\zeta_1}-1)(q^{2\zeta_2}-1)\over q^{2\zeta_0+2\zeta_2-1}}\times}{
\left( 2 q^{2\zeta_0+2\zeta_2} +2 q^{2\zeta_0+2\zeta_1+2\zeta_2}+q^{4\zeta_0+2\zeta_1+4 \zeta_2}+1\right).}$$  
We denoted by $a_{i,j,k} (r,s,t)$ the element of the algebra $a_i[r] a_j[s] a_k[t]$ where
$a_i[r]$ is $G_r, G_r^{-1},E_r$ or $1_r$, if $i=1,2,3,4$, respectively. Here, $q=e^i$.

We checked these relations for the $B_3$ vertex model, which is a four blocks model, numerically,  and indeed they hold
for various values of the heights and for general values of $q$.

\ack
It is my pleasure to thank Hans Wenzl for many discussions, encouragement  and for sending me his paper
\r\Wenzl. I am also grateful to Ida Deichaite for remarks on the manuscript and valuable
impetus. I also thank Jian-Rong Li for comments. I thank Vladimir Belavin for help with
the calculations.

\refout

\bye